\documentclass[]{article}
\usepackage[english]{babel}
\usepackage[utf8]{inputenc}
\usepackage{amsmath}
\usepackage{graphicx}
\usepackage[colorinlistoftodos]{todonotes}
\usepackage{authblk}
\usepackage{hyperref}
\usepackage{ulem}
\usepackage[margin=1in]{geometry}
\def\apj{The Astrophysical Journal}
\def\apjl{The Astrophysical Journal Letters}
\def\apjs{The Astrophysical Journal Supplement}

\def\solphys{Solar Physics}

\usepackage{natbib}
\setcitestyle{authoryear,open={(},close={)}}

\title{Machine Learning in Heliophysics and Space Weather Forecasting: A White Paper of Findings and Recommendations}
\author[1]{Gelu Nita}
\author[2,3]{Manolis Georgoulis}
\author[4]{Irina Kitiashvili}
\author[4,5]{Viacheslav Sadykov}
\author[6]{Enrico Camporeale}
\author[1]{Alexander Kosovichev}
\author[1]{Haimin Wang}
\author[1]{Vincent Oria}
\author[1]{Jason Wang}
\author[2]{Rafal Angryk}
\author[2]{Berkay Aydin}
\author[2]{Azim Ahmadzadeh}
\author[7]{Xiaoli Bai}
\author[8]{Timothy Bastian}
\author[2]{Soukaina Filali Boubrahimi}
\author[1]{Bin Chen}
\author[9]{Alisdair Davey}
\author[1]{Sheldon Fereira}
\author[1]{Gregory Fleishman}
\author[1]{Dale Gary}
\author[1]{Andrew Gerrard}
\author[10]{Gregory Hellbourg}
\author[11]{Katherine Herbert}
\author[12]{Jack Ireland}
\author[1,13]{Egor Illarionov}
\author[14,15]{Natsuha Kuroda}
\author[1]{Qin Li}
\author[1]{Chang Liu}
\author[1]{Yuexin Liu}
\author[1]{Hyomin Kim}
\author[2]{Dustin Kempton}
\author[2]{Ruizhe Ma}
\author[2]{Petrus Martens}
\author[16]{Ryan McGranaghan}
\author[17]{Edward Semones}
\author[1]{John Stefan}
\author[1]{Andrey Stejko}
\author[12,18]{Yaireska Collado-Vega}
\author[1]{Meiqi Wang}
\author[1]{Yan Xu}
\author[1]{Sijie Yu}

\affil[1]{New Jersey Institute of Technology}
\affil[2]{Georgia State University}
\affil[3]{Academy of Athens}
\affil[4]{NASA Ames Research Center}
\affil[5]{Bay Area Environmental Research Institute}
\affil[6]{University of Colorado, Boulder}
\affil[7]{Rutgers, The State University of New Jersey}
\affil[8]{National Radio Astronomy Observatory}
\affil[9]{National Solar Observatory}
\affil[10]{California Institute of Technology}
\affil[11]{Montclair State University}
\affil[12]{NASA Goddard Space Flight Center}
\affil[13]{Moscow State University}
\affil[14]{University Corporation for Atmospheric Research, Boulder}
\affil[15]{Naval Research Laboratory}
\affil[16]{ASTRA LLC}
\affil[17]{NASA Johnson Space Center}
\affil[18]{Community Coordinated Modeling Center}

\date{\today}
\begin{document}
\maketitle

\begin{abstract}
The authors of this white paper met on 16-17 January 2020 at the {\it New Jersey Institute of Technology}, Newark, NJ, for a 2-day workshop that brought together a group of heliophysicists, data providers, expert modelers, and computer/data scientists. Their objective was to discuss critical developments and prospects of the application of machine and/or deep learning techniques for data analysis, modeling and forecasting in Heliophysics, and to shape a strategy for further developments in the field. The workshop combined a set of plenary sessions featuring invited introductory talks interleaved with a set of open discussion sessions. The outcome of the discussion is encapsulated in this white paper that also features a top-level list of recommendations agreed by participants.
\end{abstract}

\section{Executive Summary}

For the economy of the discussion, we provide below the list of recommendations borne out of the workshop’s  deliberations. These  recommendations  are  intended  for  the  sponsor of  this,  namely  the  National  Science Foundation (NSF) and, potentially, other interested agencies.  We believe that this white paper (and possibly other related papers) should be distributed to the wider community, soliciting views and opinions on which the agency(-ies) should reflect on via expert teams.

In Section 2 we introduce the Research  Coordination Network that organized the workshop, while in the following sections we summarize the discussion that led to these recommendations.
\vspace{7pt}\\
\noindent Our agreed-on top-level recommendations are the following:

\begin{itemize}

\item High-quality benchmark datasets should be developed in all areas of heliophysics and space weather forecasting. Diverse models designed to tackle certain problems should be compared against these benchmarks.

\item In terms of models, laboratory plasma experiments should be combined with physical models of heliophysics processes. They provide an essential link to kinetic plasma processes whose microscopic nature precludes direct observation.

\item Machine learning (ML) is not a competitor of physical modeling, but an essential complementary asset that has been enabled by our exponentially increasing computing capability. ML is necessary because of its unmatched ability to handle the present and future Big Data environment.

\item ML or physical modeling in heliophysics and space weather forecasting entails three core quality-assurance processes that must be addressed in tandem: data, model and performance verification.

\item A standard procedure on ML applications to heliophysics and space weather forecasting should be to assign expert teams to formulate ideas and then encode these ideas on a planned course of action with certain timelines and milestones.

\item It is imperative to establish an interdisciplinary heliophysics \& ML community. This community should be served by regular workshops, such as the ML-Helio workshop series.

\item ML will be instrumental in tackling the NSF's grand-challenge projects related to heliophysics, in general, and space weather forecasting, in particular.

\item Data sets and source codes of physics-based / ML models developed to handle these data should become openly available to the community. We realize that there are challenges in this respect but encourage concerted work toward raising obstacles and ultimately achieving this goal.

\item Besides NSF, potential stakeholders in this effort to assess the benefits of ML in heliophysics and space weather forecasting include NASA, NOAA and AFOSR, but most importantly the wider, interdisciplinary heliophysics and ML community. This vibrant, diverse community should remain involved and well-informed by the highest possible communication channels.

\end{itemize}

\section{EarthCube Research Coordination Network: Towards Integration of Heliophysics Data, Modeling, and Analysis Tools}

The field of solar physics comprises instrumentation, observations of the Sun, theory, modeling and data analysis tools. This methodological arsenal, unmatched by any other standards in astronomy and astrophysics, has been developed because the Sun is the astronomical laboratory closest to us. Global, concerted efforts strive to provide the science community and the public with quantitative diagnostics of the solar structure and activity from gamma rays to radio waves, within timescales ranging from microseconds to centuries. Detailed knowledge of solar dynamics, from core to corona, is critical to connecting the Sun to the heliosphere, and hence to understanding and predicting the impacts of the Sun on Earth.
The discipline of Space Weather Forecasting aims to understand and quantify the impact of the Sun to the heliosphere, in general, and to Earth, in particular. To make progress in this front, we need to (i) understand the physical mechanisms behind the observed complex, multi-scale processes in the Sun and (ii) develop a practical and accessible cyberinfrastructure aiming to collect, access, analyze, share and visualize all forms of data from ground- and space-based  observatories, as well as from data-driven numerical simulations and modeling. These needs and goals require an osmosis between classical solar physics, Big Data and computer science.

To help achieve these goals, a Research Coordination Network (RCN) has been proposed and supported by NSF through the EarthCube (EC) AGS-1743321 grant. The Network is intended to foster new collaborations between heliophysics and computer science. This \textbf{EC} RCN, named ``Towards \textbf{I}ntegration of \textbf{H}eliophysics \textbf{D}ata, \textbf{M}odeling, and \textbf{A}nalysis \textbf{T}ools" (\textbf{@HDMIEC}), is expected to result in ideas and concrete plans for developing building blocks supporting the goals of the EarthCube initiative.

The @HDMIEC RCN actively encourages involvement of a diverse pool of community members and provides an organized virtual structure that promotes collaboration on open source knowledge discovery. This collaboration is intended to enhance the science return of individual research projects run by the RCN participants.

To initiate the process, the Steering Committee of the @HDMIEC RCN organized a kickoff workshop that was held 14-16 November 2018 at NJIT, which was attended by 43 participants and featured 19 invited presentations, including 7 presentations by representatives of other domains of geophysics funded by the EarthCube program.

The main outcome of this workshop was the establishment of four RCN working groups:

\begin{itemize}
\item WG1: ``Benchmark Datasets for Reproducible Data-driven Science in Solar Physics” RCN Working Group, founded and led by Rafal A. Angryk (GSU/Computer Science), Berkay Aydin (GSU/Computer Science), Manolis K. Georgoulis (GSU/Physics and Astronomy), and Petrus C. Martens (GSU/Physics and Astronomy)
\item WG2: ``Cross-Analysis and Validation of the Heliophysics Models and Observations” RCN Working Group founded and led by Irina Kitiashvili (NASA Ames/BAERI), Serena Criscuoli (NSO), Viacheslav Sadykov (NASA Ames/BAERI), and Alexander Kosovichev (NJIT)

\item WG3: ``Uniform Semantics and Syntax of Solar Observations and Events” RCN Working Group founded and led by Neal Hurlburt (LMSAL), Gelu M. Nita (NJT), Vincent Oria (NJIT)

 \item WG4: ``Progress toward Heliophysics data science through interdisciplinary activities," founded by Ryan McGranaghan (ASTRA, LLC) and Asti Bhat (SRI International).
\end{itemize}

At the initiative of WG1 and with support from the other working groups, the @HDMIEC RCN organized a focused 2-day workshop entitled ``Machine Learning in Heliophysics and Space Weather Forecasting: Advances, Perspectives and Synergies," the outcome of which is presented in this white paper.

\section{Current Standing of Machine Learning Applications in Heliophysics and Space Weather Forecasting}

In recent years we have witnessed a nearly explosive increase of machine learning (ML) applications in heliophysics, predominantly targeted toward space weather forecasting and the response of geospace (from the magnetosphere to Earth's ionosphere and even closer, toward Earth's surface). The benefits, if any, of this cross-pollination have yet to be proven. Rather than providing an exhaustive account of the recent literature, that already enjoys a prominent synopsis in the recent book \textit{Machine Learning Techniques for Space Weather}\citep[Eds.][]{Camporeale2018}, we attempt here a narrative of the main points of the discussion session held at the end of the @HDMIEC RCN workshop's first day. The discussion was formatted as a question-and-answer (Q\&A) session to-and-by experts, with questions aiming to delineate the current state-of-the-art of ML methods and their applications in heliophysics. The specific questions and answers are outlined in detail in the Appendix.

In inquiring about the community's level of understanding of ML capabilities and limitations, about half of the attendees responded  that the heliophysics community does not have even a fair understanding of them, or the current state-of-the-art. The remaining half split between the opposite view or remained neutral. The opinion that even 'traditional' journals in the field have trouble dealing with interdisciplinary manuscripts combining ML and heliophysics was voiced. There appeared to be a consensus that the synergy between ML and heliophysics is not yet formed well enough. This was a major incentive for the current white paper, particularly in the framework of the @HDMIEC RCN.

The sub-fields in which ML methods appear to be most successful in the field of heliophysics at this time were deemed to be the solar-disk segmentation and solar flare prediction, so a  feature or pattern recognition and a forecasting problem, respectively. ML approaches were deemed more successful in their ability to navigate through the existing Big Data landscape compared to conventional, physics-based methods (e.g., magnetohydrodynamical modeling). However, a robust and stringent comparison between the two approaches is lacking. Performing this comparison was urged by several attendees.

There was no consensus regarding which areas could ML methods clearly outperform physics-based modeling. A relatively small number of attendees favored solar-flare or CME arrival-time prediction. Some felt the question was too narrow for the current, still shaping,  state-of-the-art, while others felt that the speed of empirical vs. physics-based modeling should be a target for ML methods. Speed, along with a verifiable performance for ML methods, should be key factors for consideration.

A question was whether the attendees saw any obvious disappointment over ML applications, despite their initial high promise. The most significant response was that there are cases where simple, long-standing empirical modeling works equally well, if not better. This was generally judged as attesting to the still-evolving ML state-of-the-art and the still outstanding task of better tuning ML methods to the various problems at hand.

An optimistic tone stemmed from the question whether attendees had a clear idea of combining physics-based and ML models. The majority of responders agreed or strongly agreed. A number of clarifications were provided. In particular, an opinion was to list the current heliophysics problems where physics-based modeling clearly fails and prioritize the use of ML methods on them. Another opinion was that the answer is not black-and-white, but we should nonetheless impose physical laws in the application of ML models; otherwise, we loose  touch with the problem. Interpretability of ML methods appeared to be an area of consensus, as opposed to using 'black-box' methods judged solely by their statistical performance. In this respect, performance pointers, such as skill scores, should also be carefully chosen to assist the physical understanding on a case-by-case basis, rather than just tailored on the specifics of ML methods.

A question entertained the 'provocative' view that ML methods and physics-based modeling  should not be combined, because they serve different purposes. An overwhelming majority of the responders strongly disagreed, mainly on the grounds that physics and ML are complementary, rather than competing. This understanding promotes the use of physics-based modeling to confirm results of ML methods and vice versa. A prominent example is that a number of heliophysics problems have a solid statistical element alongside physics (the solar cycle prediction, for example), where ML models can play an instrumental role in modeling and predicting the statistics that can decisively help us understand the underlying physics.

Concluding the session, a question adopted the view that the discussion on ML methods is a 'bubble' that will burst if we apply ML methods indiscriminately, even to problems in which we do not expect them to work or we do not understand why they should work. About half of the responders disagreed, with the remaining half agreeing or staying neutral. A majority, therefore, did not feel that we live in a 'bubble' of ML frenzy, but at the same time a clear opinion was voiced that explanations of why a given ML method is applied to a given problem are desirable, or even necessary.

On top of these questions, a free-format discussion highlighted an important  distinction that must be made between ML applications for space physics and those for space weather forecasting. Both space physics and space weather forecasting are integral parts of heliophysics. However, every time a ML method is applied, the community should ask what is the objective: more accurate predictions or a better physical understanding of the processes at work? Responses find it important to clarify both the ML method(s) to be used and their specific setup.

\section{Envisioned Contributions by this RCN}

The discussion during the second day of the workshop revolved around  the envisioned contributions of the @HDMIEC RCN. The following is a concise encapsulation of it, embracing the SMART principle: the outcome of the RCN should be \emph{S}pecific, \emph{M}easurable, \emph{A}ttainable, \emph{R}elevant and \emph{T}imely, linking the big picture (heliophysics, in terms of understanding and forecasting) to the finer detail (specific problems), feeding back into the big picture in a integrated, seamless and modular fashion.

Considering the more detailed picture, tackling a given problem requires a theory whose main purpose is to make testable predictions. This further constrains the data on which a theory is to be applied, the modeling that will implement the theory on the data and, finally, the verification of the models' performance. Before performance verification is enabled, however, data and model must be verified accordingly. The following sections describe these steps, in connection to the @HDMIEC RCN.

\subsection{Data Verification and Benchmark Datasets}

The @HDMIEC RCN's WG1 refers to data benchmarking. A benchmark dataset, particularly under ML applications, is an integral part of the knowledge discovery process, according to the WG1 leaders. It is a fixed testbed, or breadboard, on which new ideas, materialized into models, are tested, with an  objective to test the reproducibility, effectiveness and efficiency of the models. It is of little meaning to test different models on non-benchmarked datasets.

Benchmark datasets must be of the highest-possible quality, given data imperfections. These flaws stem from the fundamental nature of data, that are essentially imperfect readings of real physical systems, observed by imperfect instruments. Instrumental caveats are present in terms of measurement uncertainties, as well as in terms of finite spatial, temporal and spectral resolution. Data quality is typically verified by cross-checking and calibration between similar information provided by different instruments. Essentially, this process uses thresholding, as it is highly unlikely that similar information by different instruments will be identical. The similarity or 'closeness' threshold(s) extend from the theory and its physical understanding of the system, either directly or statistically, by eliminating impossible alternatives or via maximum likelihood analyses.

Prominent examples are benchmark datasets used for solar flare prediction, which are verified in terms of flare source location, magnitude and timing. These are typically verified by different observing instruments. Non-verifiable data, particularly in case they are uniquely obtained by a single existing instrument, are still used out of necessity (for example, in-situ coronal mass ejection properties at L1) but in this case differences between models cannot be solely attributed to the models' different logic, but also to their varying susceptibility to the (largely unknown) uncertainties of the data.

WG1 of the @HDMIEC RCN is mainly focused on data verification and benchmarking.

\subsection{Model verification and cross-comparison between observations and laboratory plasma experiments}

The dramatic increase in observational data flow from ground- and space-based telescopes (e.g., SDO, IRIS, Hinode, Parker Probe, BBSO, etc.), as well as from upcoming observational facilities, such as the 4-meter solar telescope DKIST, provides a unique opportunity to probe solar dynamics from interior to corona with high spatial and temporal resolution. At the same time, it is a great challenge to analyze large multi-dimensional data sets and interpret the observed highly nonlinear microscopic (plasma) and macroscopic (magnetohydrodynamical [MHD])  processes.
Recent advances in 3D radiative MHD modeling make it possible to reproduce the dynamics of subsurface magnetoconvection layers and the solar atmosphere with a hopefully high degree of realism, providing tools for a physics-based interpretation of observational data.

The ability of models to reproduce certain phenomena or conditions depends on accuracy of the radiative transfer approach, effects of turbulence, the assumed initial and boundary conditions, numerical schemes, spatial resolution, etc. Comparison  with observational data of the resulting models, as well as with the modeled spectroscopic and spectropolarimetric synthetic observables, allows researchers to perform model validation, evaluate the models'  quality, limitations, and optimize code parameters to achieve more accurate descriptions of the magnetized solar plasma. Connection of laboratory plasma experiments to modeling provides an essential link to processes on kinetic scales that cannot be resolved by the imperfect observations described above.

ML plays an increasingly important role in bridging together observations, numerical simulations, and experiments. There are several possibilities of how ML helps cross-link these different areas, including (but not limited to):
\begin{enumerate}
    \item Adoption of realistic, fully 3D MHD models and synthesized emission for training ML algorithms, further applied to observational data \citep[e.g.][]{Osborne2019,Tremblay2018};
    \item Enhancement of the computationally-demanding physics-based inversion models by their ML-based analogs \citep[e.g.][]{SainzDalda2019,Wright2019};
    \item Development of ML-based algorithms connecting observational inferences to physical models \citep[e.g.][]{Galvez2019,Kim2019}.
\end{enumerate}

Detailed physical interpretation of complex processes observed in the Sun is another challenge. Therefore, development of models in conjunction with ML analysis of multi-dimensional data sets (a Big Data environment) is part of the focus of the @HDMIEC RCN WG2.

The goal of the @HDMIEC RCN Working Group “Cross-Analysis and Validation of the Heliophysics Models, Laboratory Experiments, and Observations” is to create a platform for the interaction of observers, data analysts, and modelers to identify key points and challenges that must be addressed to advance knowledge of physical processes occurring in the Sun. WG2 activities include discussions of current challenges and new ways to efficiently resolve them, thus promoting cross-disciplinary activities and establishing collaborations.

\subsection{Performance Verification}

Following the verification of data and models is the verification of the model performance. Verification methods and related scores, skill scores and statistical metrics are not necessarily tied to the actual physical problem but to much wider categories. For prediction problems, these categories are binary (YES/NO), (multi-)categorical, or probabilistic forecasting; other categories exist for non-forecasting problems. While there is substantial literature on verification methods spanning over more than a century, the @HDMIEC RCN does not have a dedicated Working Group on performance verification. This would be a very worthy enhancement in similar future efforts.

A key task with performance verification is to choose from the available diversity of techniques the ones most relevant to a given problem or, better yet, to a class of problems in heliophysics. For prediction problems there are the above categories along with hybrid techniques cross-cutting between them. In addition, performance verification relies on a robust training and testing of ML methods that should follow certain rules for the validation to have any significance or meaning.

Defining tasks, goals and generally accepted validation practices is considered a hard, but urgently needed, action. This action is undertaken by both NASA and the European Space Agency (ESA), at least. Within ESA, best practices are actively discussed within the Space Weather segment of the Agency's Space Situational Awareness (SSA) Programme, while NASA's Community Coordinated Modeling Center (CCMC) has established Scoreboards for solar flares, coronal mass ejections and solar energetic particle event forecasts.

We reiterate, finally, that performance verification is not an exclusive trait of forecast models, but it also applies to models devoted to physical understanding. In this case the 'testable prediction' intrinsic to the theory on which models rely should involve expected patterns and behaviors to justify the validity of the models' performance. Much of the focus on performance verification in heliophysics modeling, however, has been directed toward prediction, apparently because of the immense interest in the buildup of reliable space weather forecasting capabilities.

\subsection{Takeaways from the general discussion}

During the second-day discussion it was deemed important to raise the visibility of the @HDMIEC RCN among different groups and agencies. Attendees from NASA were interested in contributing toward an increase in awareness of the RCN beyond the funding NSF.

A part of the discussion focused on the imminent Daniel K. Inoue Solar Telescope (DKIST) and on whether its future data should be acquired by the RCN or its potential successor. An attendee from the DKIST team confirmed that significant infrastructure should be in place for collecting the DKIST data but its importance to understanding, primarily, and forecasting, secondarily, of the quiescent and eruptive dynamics in the low solar atmosphere would be high.

Discussion covered future infrastructure needs to accommodate the expected substantial data volumes. It was mentioned that this should be task- and problem-specific, as attempting to assemble heliophysics data indiscriminately would be impractical. Nonetheless, there are significant efforts currently underway at NASA/Ames and Stanford University, that also pay due attention to data quality.

The target audience of the present white paper was also discussed. The white paper is intended primarily for the funding Agency (NSF) as a means to keep the momentum and, perhaps, attract future projects. This, however, does not imply that community recommendations should be left behind. One of these recommendations was that we need a standard procedure toward ML application: ideas formulated by a team of experts and then a planned course of action based on these ideas. NASA, in particular, currently engages in the development of ML (for example, via its Frontier Development Lab) and is, therefore, in need of benchmark datasets.

Community involvement should be solicited by widely distributing the white paper, requesting comments, and then reflecting on these comments in the text. Previous available, well-organized white papers could also be consulted for consistency. The need for the establishment of an interdisciplinary heliophysics / ML community was stressed. This community could be served with regular workshops. In fact, this is already implemented by the \textit{Machine Learning in Heliophysics} (ML-Helio) series of workshops \citep[the next being planned for Summer 2021;][]{camporeale2020ml}.  In terms of the funding Agency, it was agreed that ML will be indispensable in tackling NSF's grand challenge projects related to space weather.

A key factor for the success of this common endeavor is openness of data and applicable codes, extending all the way to source codes. A concern was voiced as to how would one be compelled to share one's source code from a project that is not publicly- or self-funded. This is a valid point, indeed, but a number of attendees expressed the view that with more and more source codes becoming public, community members with proprietary codes will eventually be compelled to share, as well. This said, it currently lies entirely at the discretion of a developer / modeler to share their intellectual products under public, no-liability licensing. On whether codes should be available indiscriminately for the entire community or just for registered members, a number of attendees felt this this falls on the licensing policies and guidelines followed by the funding agencies.

\section{Potential Stakeholders and Synergies}

Which stakeholder(s) would be potentially interested in a synergy with the @HDMIEC RCN or its future successor? An attempt to answer this question focuses on the main pylons of envisioned @HDMIEC contributions, namely, (benchmark) data, models, and performance verification.

\textbf{Data:} There is significant expertise in data benchmarking by the RCN's WG1, with participation from the GSU's Data Mining Lab. In addition, NJIT's Institutes for Space Weather Sciences (ISWS) and Data Sciences (IDS) can  contribute meaningfully. The wider community could also weigh in heavily by contributing more datasets appropriate for ML treatment and tailored to heliophysics problems.

\textbf{Models:} There are a number of models assembled by the RCN's WG2. While NASA's CCMC does not release the source codes from its model inventory (this is part of the agreement with model contributors to the CCMC) a synergy with the CCMC would be desirable and was encouraged during the workshop. NOAA and the AFOSR could also be interested, in principle, to provide models and relevant expertise. Finally, the wider community could have a distinct role and contribution.

\textbf{Performance verification:} The RCN could tap into the NSF supporting and networking resources to link with parts of the science community with demonstrable  expertise. This would lead to a buildup of this expertise within the RCN or its potential successor. There could also be an strategic collaboration between the RCN and CCMC with its scoreboards, as the discussion on best practices in performance verification is dominant within them. NOAA also has a documented, keen interest in performance verification, along with AFOSR. These agencies could be contacted for additional feedback. Finally, once again, the wider community could be consulted via targeted communications with individuals and groups generally accepted as experts in these tasks.

One notices the common element for interaction, contact or consultation in all above areas of the RCN's contributions: this is the wider science community. It it seen as a need, therefore, to constantly pursue an interaction with the community, starting from this white paper.

\section*{Appendix: Outcome of the Q\&A sessions during the first day of the @HDMIEC Workshop}
The questionnaire below and the responses provided reflect only the collective views of the attendees of this workshop. We provide them in this light and understanding, although we feel that a similar set of questions addressed by the wider heliophysics community would be well warranted.
\begin{figure}[h]
    \centering
    \includegraphics[width=0.85 \textwidth]{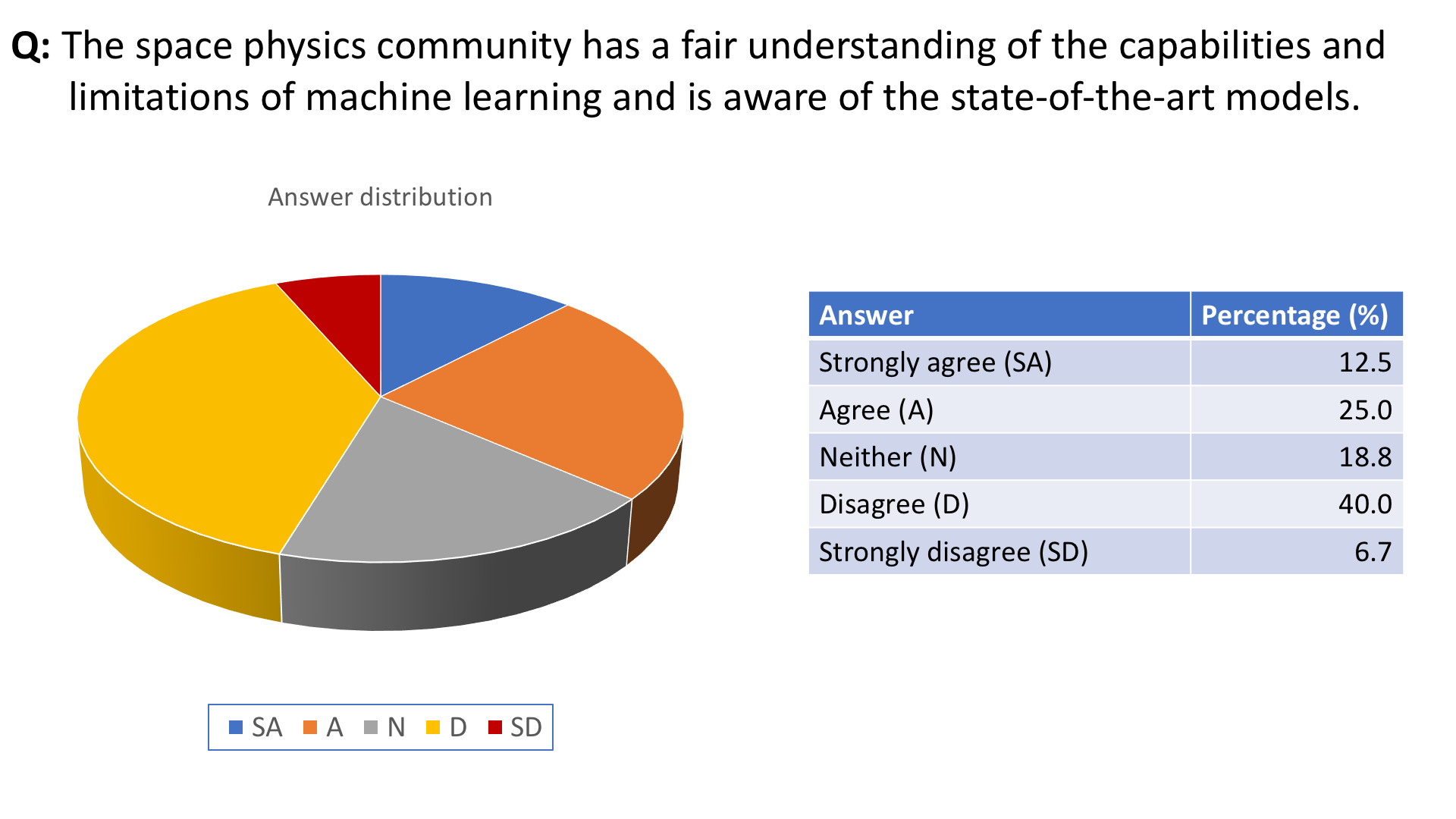}
    \includegraphics[width=0.85 \textwidth]{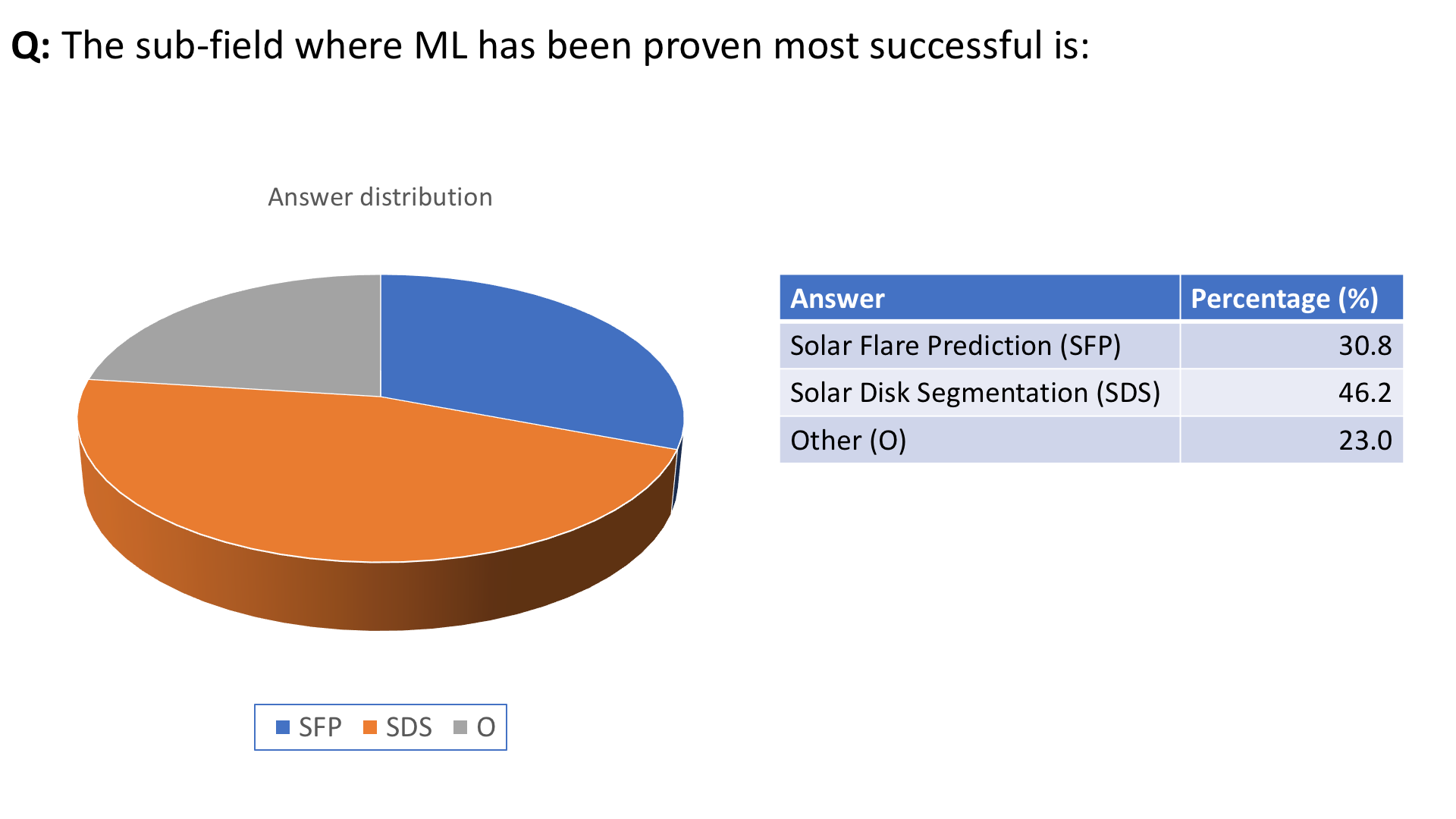}
\end{figure}
\begin{figure}
    \centering
    \includegraphics[width=0.85 \textwidth]{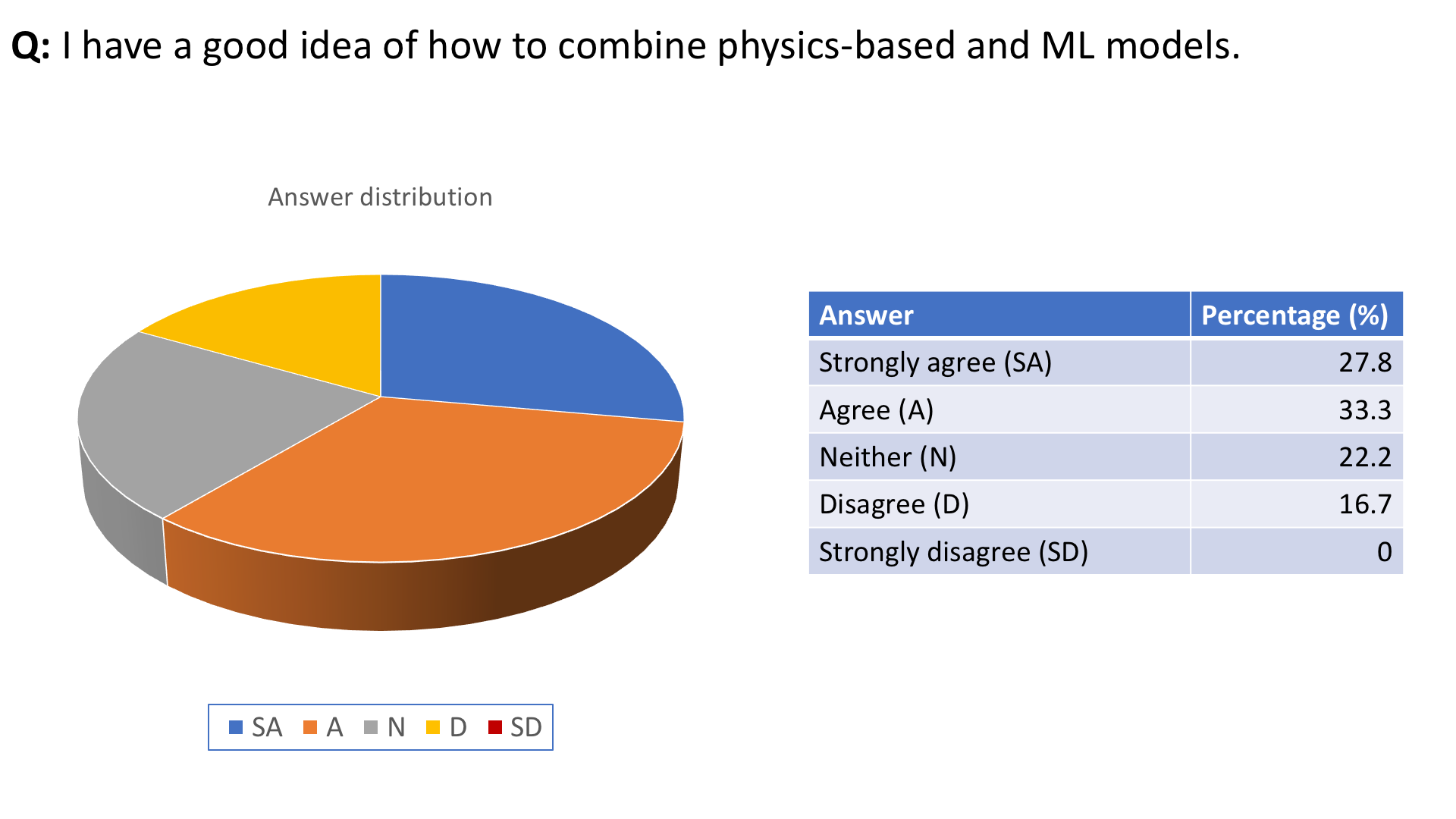}
    \includegraphics[width=0.85 \textwidth]{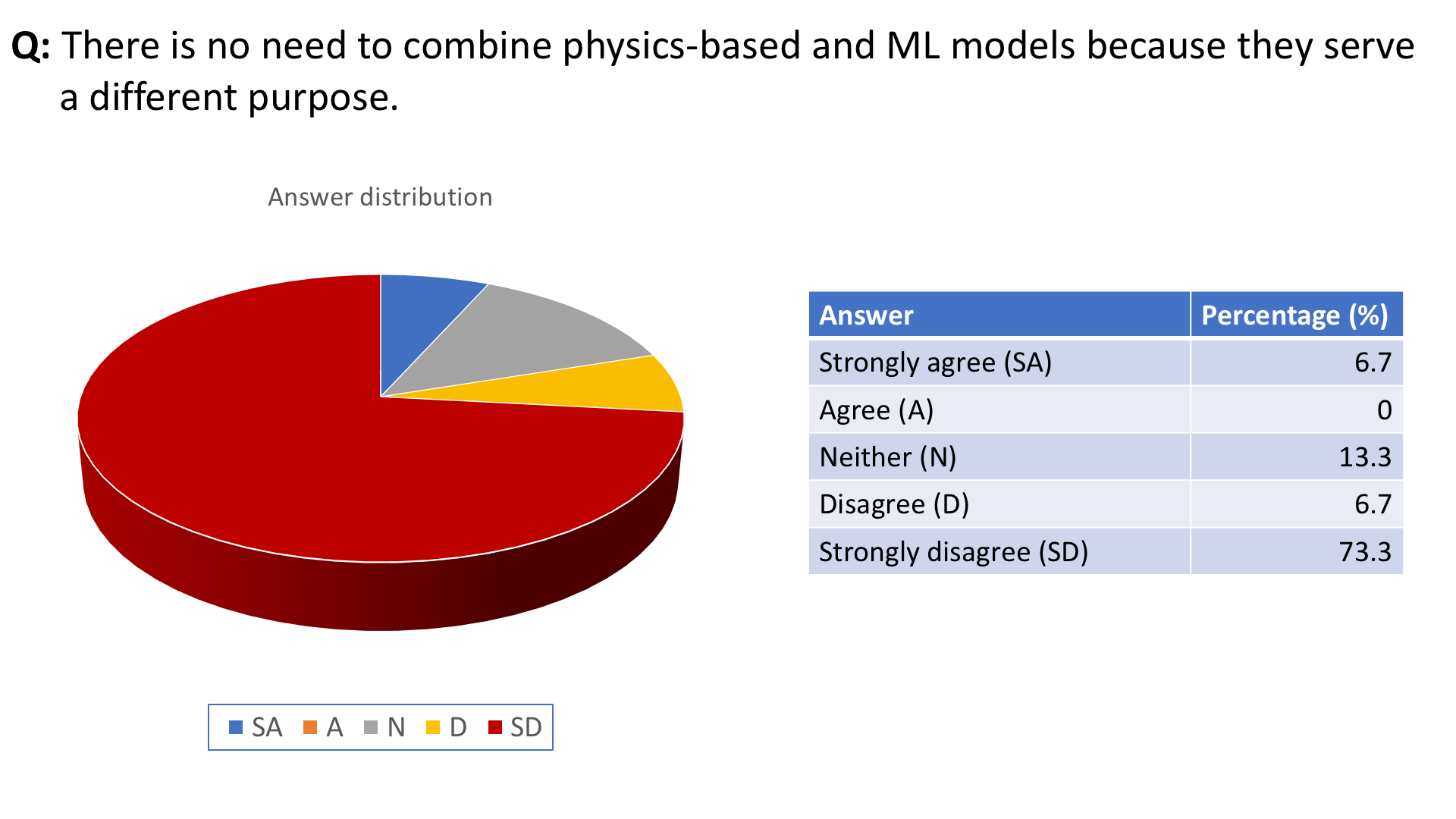}
\end{figure}
\begin{figure}
    \centering
    \includegraphics[width=0.85 \textwidth]{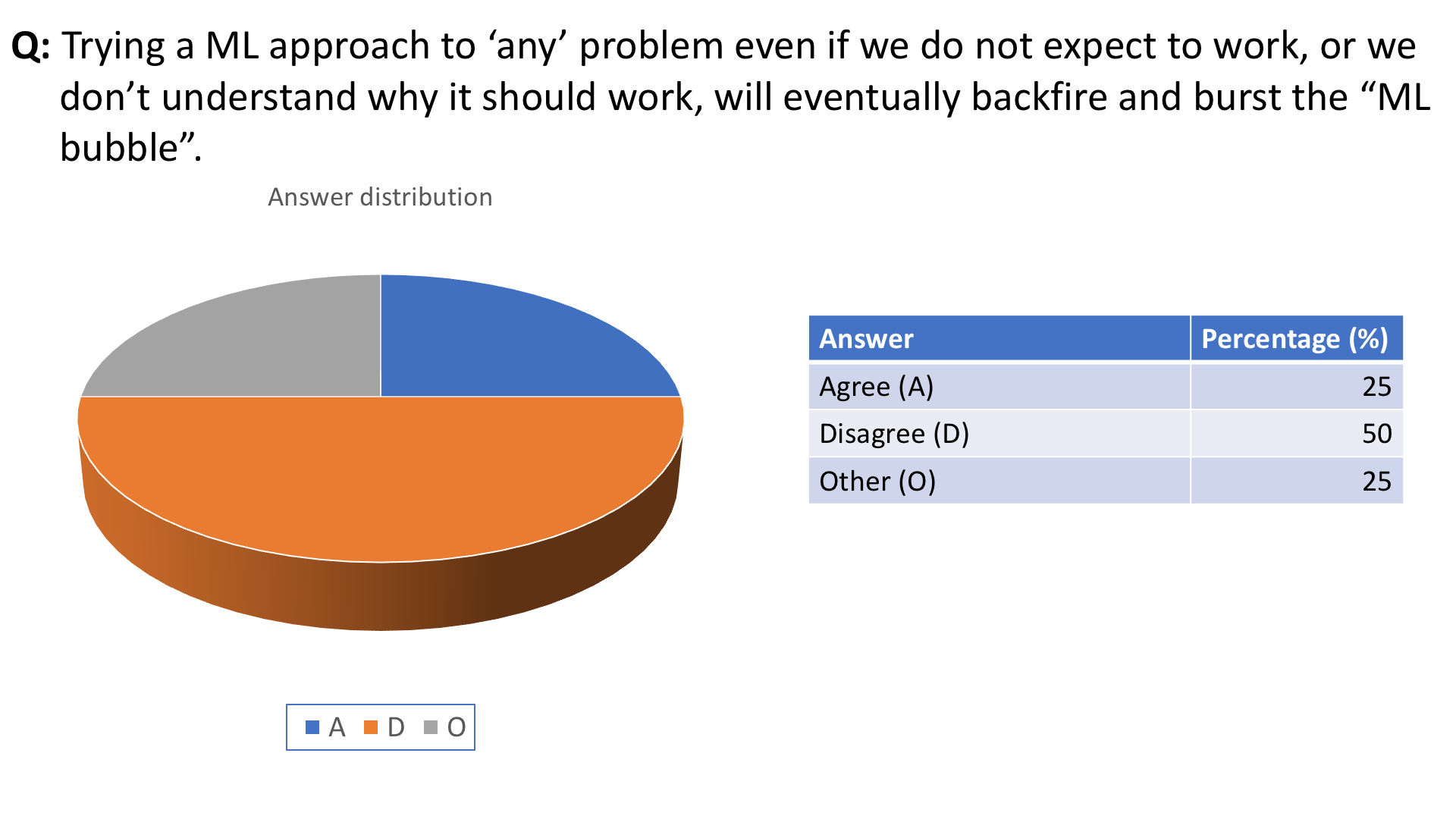}
    \includegraphics[width=0.85 \textwidth]{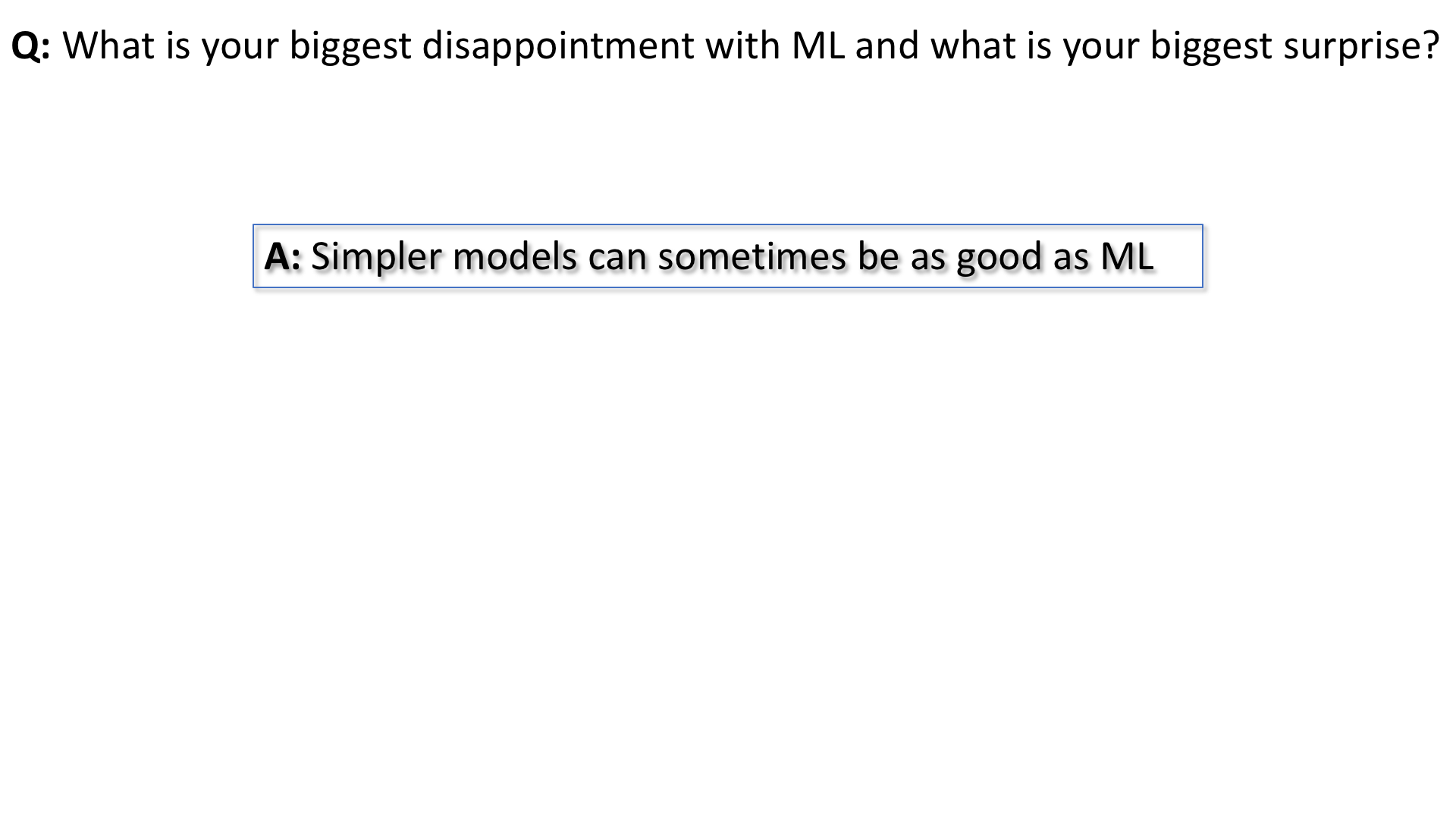}
\end{figure}
\section*{Acknowledgments}
This material is based upon work supported by the National Science Foundation under Grant AGS-1743321. Any opinions, findings, and conclusions or recommendations expressed in this material are those of the authors and do not necessarily reflect the views of the National Science Foundation.

\end{document}